\documentclass[twocolumn,superscriptaddress,prb,aps,preprintnumbers,nofootinbib,longbibliography]{revtex4-2}
\usepackage{amsmath,amssymb,graphicx,color}
\usepackage[T1]{fontenc}
\usepackage[colorlinks=true, citecolor={blue!80!black}, urlcolor={blue!50!black}, linkcolor = {blue!80!black}]{hyperref}
\usepackage{comment}

\usepackage{xcolor}

\newcommand{\hamilt}{\hat{\mathcal{H}}}
\newcommand{\spin}{\hat{S}}
\newcommand{\spop}{\hat{\mathbf{S}}}

\newcommand{\aver}[1]{\left\langle #1 \right\rangle}

\newcommand{\mo}{\mu_{0}}
\newcommand{\mub}{\mu_\mathrm{B}}
\newcommand{\kb}{k_\mathrm{B}}

\usepackage{pdfpages} 
\usepackage{pgffor} 
\makeatletter
\AtBeginDocument{\let\LS@rot\@undefined}
\makeatother
\pdfximage{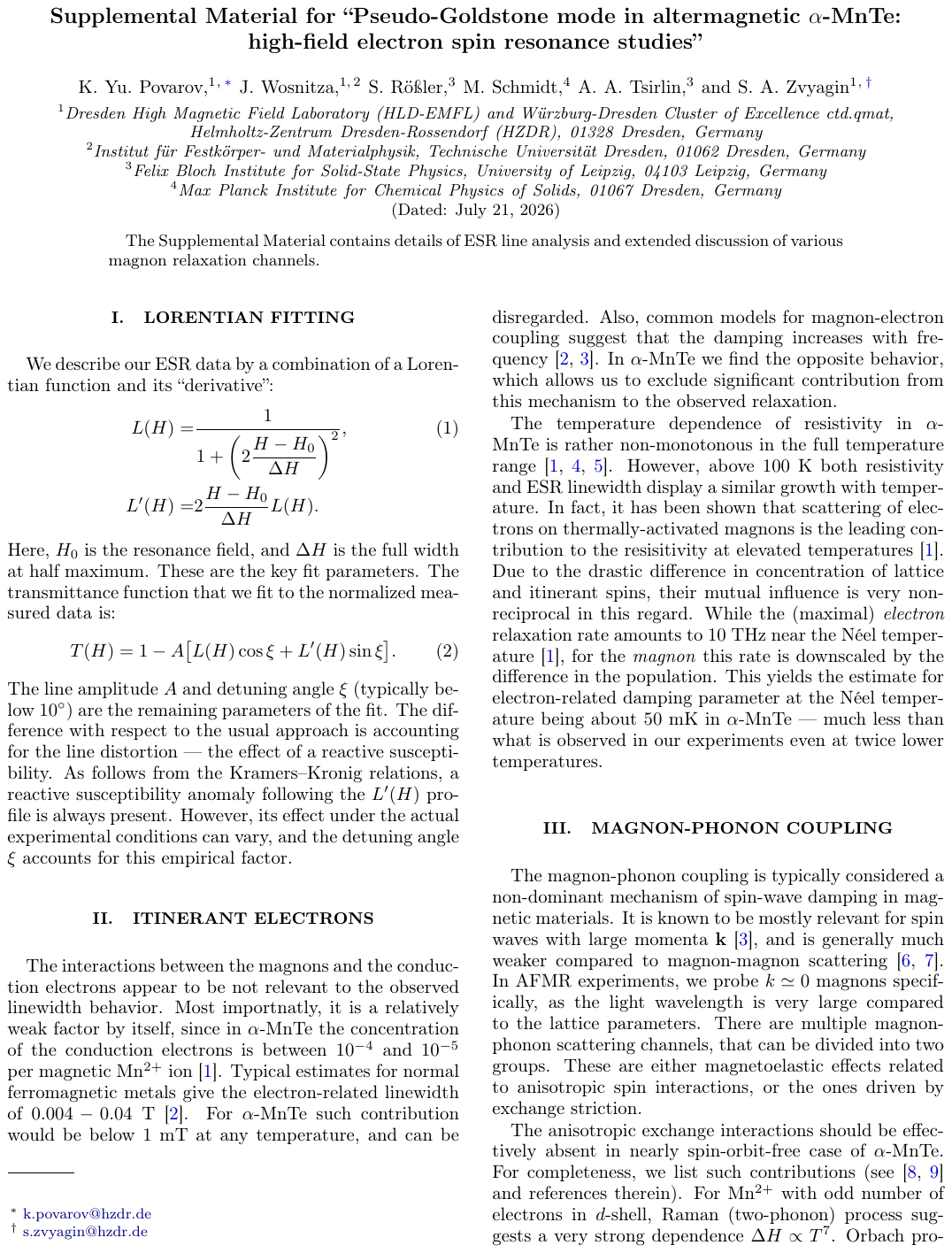}
\def\numbersupplementpages{\the\pdflastximagepages}

\begin{document}
\title{Pseudo-Goldstone mode in altermagnetic $\alpha$-MnTe: high-field electron~spin~resonance~studies}

\author{K.~Yu.~Povarov}
\email{k.povarov@hzdr.de}
\affiliation{Dresden High Magnetic Field Laboratory (HLD-EMFL) and W\"urzburg-Dresden Cluster of Excellence ctd.qmat, Helmholtz-Zentrum Dresden-Rossendorf (HZDR), 01328 Dresden, Germany}

\author{J.~Wosnitza}
\affiliation{Dresden High Magnetic Field Laboratory (HLD-EMFL) and W\"urzburg-Dresden Cluster of Excellence ctd.qmat, Helmholtz-Zentrum Dresden-Rossendorf (HZDR), 01328 Dresden, Germany}
\affiliation{Institut f\"ur Festk\"orper- und Materialphysik, Technische Universit\"at Dresden, 01062 Dresden, Germany}

\author{S.~R\"{o}\ss ler}
\affiliation{Felix Bloch Institute for Solid-State Physics, University of Leipzig, 04103 Leipzig, Germany}

\author{M.~Schmidt}
\affiliation{Max Planck Institute for Chemical Physics of Solids, 01067 Dresden, Germany}

\author{A.~A.~Tsirlin}
\affiliation{Felix Bloch Institute for Solid-State Physics, University of Leipzig, 04103 Leipzig, Germany}

\author{S.~A.~Zvyagin}
\email{s.zvyagin@hzdr.de}
\affiliation{Dresden High Magnetic Field Laboratory (HLD-EMFL) and W\"urzburg-Dresden Cluster of Excellence ctd.qmat, Helmholtz-Zentrum Dresden-Rossendorf (HZDR), 01328 Dresden, Germany}

\begin{abstract}
We report multi-frequency electron spin resonance spectroscopy studies of $\alpha$-MnTe in magnetic fields up to $16$~T, applied along the easy anisotropy axis. At temperatures below $T_\mathrm{N} = 310$~K, we observe a single resonance line corresponding to the pseudo-Goldstone mode of the antiferromagnetic resonance (AFMR). This mode exhibits the isotropic behavior with $g_\mathrm{eff}=2.01$, consistent with a complete quench of the orbital angular momenta for Mn$^{2+}$ ions. At low temperatures, the resonance mode is remarkably narrow ($\sim50$~mT for the full width at the half-maximum at $5$~K). The AFMR mode exhibits substantial broadening with increasing temperature, which can be understood in terms of the magnon-magnon scattering.
\end{abstract}
\date{\today}
\maketitle

\textit{Introduction.}
The recently introduced concept of altermagnetism~\cite{SmejkalSinovaJungwirth_PRX_2022_Altermagnetism} has sparked a great interest in the solid-state physics community~\cite{Smejkal_PRX_2022_Altermagnetism,BaiFeng_AdvFunctMaterials_2024_AltermagnetismExploring,Wei_ACSOrgInorgAu_2024_AltermagChemistryDesign,Fender_JAmChemSoc_2025_AltermagnetismChemicalPerspective}. At first glance, altermagnets are similar to antiferromagnets: the localized magnetic moments are arranged in several sublattices, such that the net magnetic moment of the sample is 0. In conventional antiferromagnets, the translation or inversion symmetry connects the opposite-spin sublattices~\cite{Bertaut_ActaCryst_1968_RepresentationAnalysis,Brown_PhysBC_1986_MagneticStructure,Borovik_2006_XTablesMagnetism}. However, in the case of altermagnets rotational or mirror transformation is necessary. The chase for finding new altermagnetic materials is ongoing~\cite{BaiFeng_AdvFunctMaterials_2024_AltermagnetismExploring,Wei_ACSOrgInorgAu_2024_AltermagChemistryDesign,Fender_JAmChemSoc_2025_AltermagnetismChemicalPerspective}. Some compounds, such as CrSb~\cite{DingJiang_PRL_2024_CrSbbandsplit,Reimers_NatCommun_2024_CrSbbandsplit}, or CoNb$_4$Se$_8$~\cite{DaleAshour_arXiv_2024_CoNb4Se8split,SakhyaMondal_arXiv_2025_CoNb4Se8bands} have already been identified as belonging to this class.

The target material of the present study, $\alpha$-MnTe, has been known as an antiferromagnet since the 1950s~\cite{Uchida_JPSJ_1956_MnTeBasics,Komatsubara_JPSJ_1963_MnTeMagnet,Szuszkiewicz_PhysStatSol_2005_MnTeNeutrons,SzuszkiewiczDynowska_PRB_2006_MnTeNeutrons,KriegnerReichlova_PRB_2017_MnTeVarious}. However, it became a subject of very intense research just recently once  it was proposed as a possible altermagnet~\cite{Mazin_PRB_2023_MnTealterm}. This altermagnetism case is especially exciting, since the Mn$^{2+}$ magnetic ions have quenched orbital momentum $L=0$.

\begin{figure}
 \includegraphics[width=0.48\textwidth]{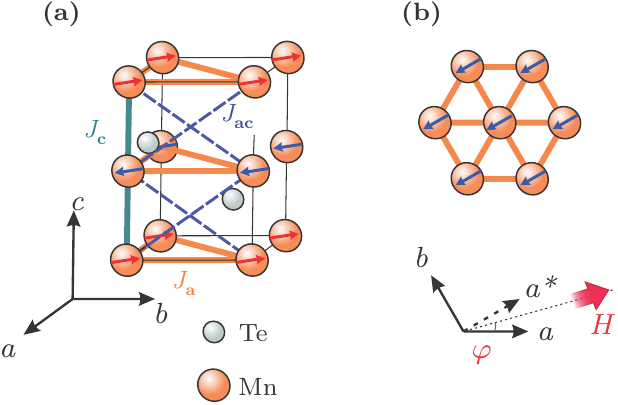}
 \caption{(a) Schematic view of the crystal structure of $\alpha$-MnTe. Big orange spheres represent Mn$^{2+}$ magnetic ions, small gray spheres correspond to Te$^{2-}$ ions. Cyan, orange, and blue lines indicate the leading exchange interactions $J_{c}$, $J_{a}$, and $J_{ac}$, correspondingly (see text for more details). Spin directions, according to the neutron diffraction data~\cite{KriegnerReichlova_PRB_2017_MnTeVarious}, are shown by red and blue arrows for different sublattices, respectively. The unit cell is shown by thin black lines.  (b) Arrangement of Mn ions and their magnetic moments in one of the $(0,~0,~1)$ layers. The direction of the external magnetic field in this plane is given by the angle $\varphi$ with respect to the $a$ direction. }\label{FIG:xtal}
\end{figure}

In $\alpha$-MnTe, the magnetic ions with $S=5/2$ are arranged in triangular-lattice layers. The material orders antiferromagnetically below $T_\mathrm{N}=310$~K~\cite{Szuszkiewicz_PhysStatSol_2005_MnTeNeutrons,Uchida_JPSJ_1956_MnTeBasics}. The magnetic structure can be described by the $\mathbf{Q}=(0,~0,~0)$ propagation vector, with spins coaligned [$4.75(15)\mub$ per Mn$^{2+}$ ion] within the $(0,0,1)$ planes along the direction that alternates between the layers, as illustrated in Fig.~\ref{FIG:xtal}~\cite{Szuszkiewicz_PhysStatSol_2005_MnTeNeutrons,KriegnerReichlova_PRB_2017_MnTeVarious}. Unusual properties of $\alpha$-MnTe are manifested in numerous experiments. Angle-resolved photoemission confirms electron band splitting, suggested to take place in altermagnetic materials~\cite{LeeLee_PRL_2024_MnTeARPES,OsumiSouma_PRB_2024_MnTeARPES,KrempaskySmejkal_Nature_2024_MnTeARPES}. Inelastic neutron scattering reveals splitting of magnon bands as well~\cite{LiuOzekiAsai_PRL_2024_MnTeInelastic}. The altermagnetism of $\alpha$-MnTe also reveals itself in x-ray dichroism~\cite{Amin_Nature_2024_MnTeImage,HarikiDal_PRL_2024_MnTeXRCD} and the anomalous Hall effect~\cite{KluczykGas_PRB_2024_MnTeAHE}. It is also considered a promising material for spintronic memory applications~\cite{GonzalezBetancourt_npjSpintronics_2024_MnTeMR,KriegnerReichlova_PRB_2017_MnTeVarious,KriegnerVyborny_NatCommun_2016_AMRinMnTe}.

Magnetic properties of $\alpha$-MnTe can be described employing the effective spin Hamiltonian:
\begin{equation}
 \hamilt=\sum\limits_{\aver{i,j}} J_{ij}\spop_{i}\spop_{j}+D\sum\limits_{i} (\spin^{z}_{i})^2,
 \label{EQ:GenHamilt}
\end{equation}
with $ij$ spin pairs summed only once and the $z$ direction corresponding to the $c$ crystallographic axis.
The most prominent terms are the Heisenberg exchange interactions $J_{a}/\kb=-1.4$~K, $J_{c}/\kb=46.3$~K, and  $J_{ac}/\kb=5.5$~K~\cite{LiuOzekiAsai_PRL_2024_MnTeInelastic}. In addition, a small effective easy-plane anisotropy constant $D/\kb=0.5(1)$~K is present in the Hamiltonian. The bonds are also schematically shown in Fig.~\ref{FIG:xtal}(a).

The zero-field spectrum of magnetic excitations in $\alpha$-MnTe has been investigated by means of the inelastic neutron spectroscopy ~\cite{SzuszkiewiczDynowska_PRB_2006_MnTeNeutrons,LiuOzekiAsai_PRL_2024_MnTeInelastic}. An energy gap of about $850$~GHz ($40$~K) has been observed in the spin-wave spectrum at $\mathbf{Q}=(0,~0,~0)$. The temperature and field evolution of the gap has been studied by means of far-infrared spectroscopy~\cite{Dzian_PRB_2025_MnTeAFMR}, with field applied along the hard anisotropy axis ($c$ direction of the crystal).

The lack of information on the low-energy spin dynamics in applied magnetic field has motivated us to investigate $\alpha$-MnTe by means of electron spin resonance (ESR) spectroscopy. For the first time, we have experimentally tracked the antiferromagnetic resonance (AFMR) mode, corresponding to a pseudo-Goldstone spin wave at the zone center, in a broad range of magnetic fields [within the magnetic anisotropy easy plane $(0,0,1)$] and temperatures. The most important finding is the $h\nu/\kb T$ scaling behavior of the AFMR linewidth, suggesting that the resonance damping is related to magnon-magnon scattering.

\textit{Experiment.}
 In our study, we have used high-quality  single-crystalline samples of $\alpha$-MnTe ($P6_{3}/mmc$, space group 194, $a=4.15$~\AA, $c=6.71$~\AA~\cite{Kunitomi_JPhysFrance_1964_DiffractionMnTe}) grown in a two-stage process, based on Ref.~\cite{deMeloLeccabue_JCrystGrowth_1991_MnTe}.
First, MnTe was synthesized by a direct reaction of the elements in an equimolar ratio of Mn (pieces, Alfa Aesar 99.9998\%, powdered directly before use) and Te (powder, Alfa Aesar 99.999\%) with the addition of iodine (Alfa Aesar 99.998\%) at 500~$^\circ$C in evacuated quartz glass tubes over a period of 10 days. Subsequently, starting from the microcrystalline powder synthesized, crystals of $\alpha$-MnTe were grown by chemical transport in a temperature gradient from 700~$^\circ$C (source) to 650~$^\circ$C (sink) with the addition of iodine (Alfa Aesar 99.998\%) $1.5$~mg/ml as a transport agent. The crystals have a thin platelet shape (typical size $5\times5\times0.1$~mm), with the surface perpendicular to the $c$ direction. The sample orientation was checked by Laue x-ray diffraction.

For the ESR experiments, we employed a spectrometer (similar to the one described in Ref.~\cite{Zvyagin_PhysB_2004_ESRinCuGeO3}), equipped with a $16$~T superconducting magnet. We used VDI microwave-chain sources (product of Virginia Diodes, Inc., USA) to generate radiation in the frequency range of c.a. $30-400$~GHz, and a hot-electron n-InSb bolometer (product of QMC Instruments Ltd., UK), operated at $4.2$~K,  as a THz detector. For the frequencies below $50$~GHz we employed a microwave vector network analyser (MVNA, production of AB Millimeter, France).
In our experiments, we applied magnetic field perpendicular to the $c$ axis of the crystal, using a sample holder in the Voigt configuration.

\begin{figure}
 \includegraphics[width=0.48\textwidth]{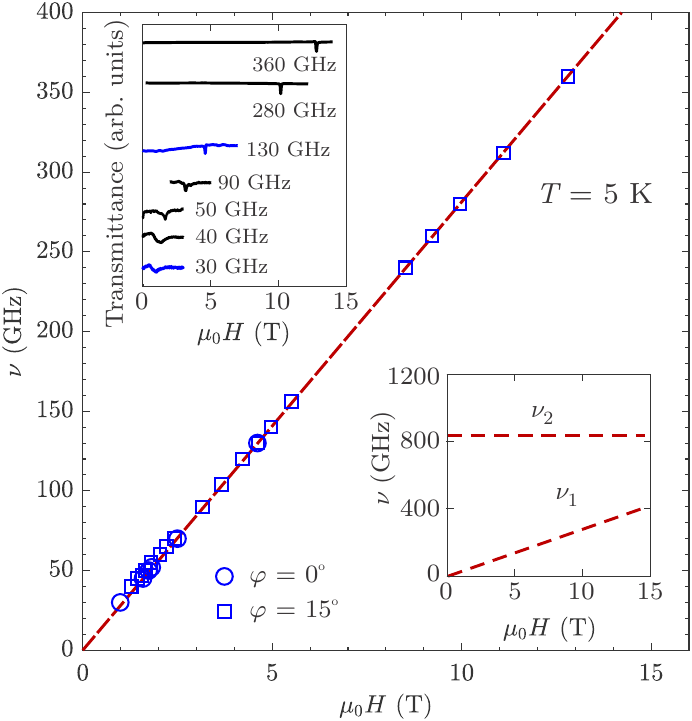}
 \caption{Main panel: Frequency-field diagram of low-energy magnetic excitations in $\alpha$-MnTe at $T=5$~K, $H\perp c$ (field within the easy anisotropy plane). Points [circles for $\varphi=0^\circ$ and squares for $\varphi=15^\circ$ in-plane magnetic field directions as defined in Fig.~\ref{FIG:xtal}(b)] are the resonance-line centers; the dashed line corresponds to $g=2.01$.
Upper inset: Exemplary ESR spectra; black lines correspond to $\varphi=15^\circ$ and blue lines to $\varphi=0^\circ$. Lower inset: Schematic frequency-field diagram of $\alpha$-MnTe in magnetic field, applied within the magnetic anisotropy easy plane, according to Eq.~(\ref{EQ:fHtrans}).}\label{FIG:FvsH}
\end{figure}

\textit{Results.}
At low temperatures, in magnetic fields applied within the $(0,0,1)$ plane, we observe a single resonance line. Some exemplary data (radiation transmittance as function of the magnetic field, $T=5$~K) are displayed in the upper inset of Fig.~\ref{FIG:FvsH}. For two non-equivalent magnetic-field directions in the $(0,0,1)$ plane, the positions of the resonance fields fall onto the same linear frequency-field dependence, as demonstrated in the main panel of Fig.~\ref{FIG:FvsH}. This linear dependence $h\nu=g_{\perp}\mub\mo H$ (where $\mub$ is the Bohr magneton and $\mo$ is the vacuum magnetic permeability) corresponds to the effective in-plane $g$-factor $g_{\perp}=2.01$. Such frequency-field relation is what one would expect for the AFMR mode of a uniaxial easy-plane antiferromagnet in magnetic field applied within the plane. The frequency-field dependences of magnetic excitations for this orientation of the applied magnetic field (for $H\ll H_\mathrm{sat}$, where $H_\mathrm{sat}$ is the magnetization saturation field) can be described
 by~\cite{GurevichMelkov_1996_Wavesbook}:

\begin{equation}\label{EQ:fHtrans}
 h\nu_{1}=g_{\perp}\mo\mub H,~h\nu_{2}=\Delta,
\end{equation}
where $\Delta$ is the zero-field AFMR gap.  The corresponding frequency-field diagram is schematically illustrated in the bottom inset of Fig.~\ref{FIG:FvsH}. In ESR experiments with field-swept technique, only the mode $\nu_1$ is observable.

\begin{figure}
 \includegraphics[width=0.45\textwidth]{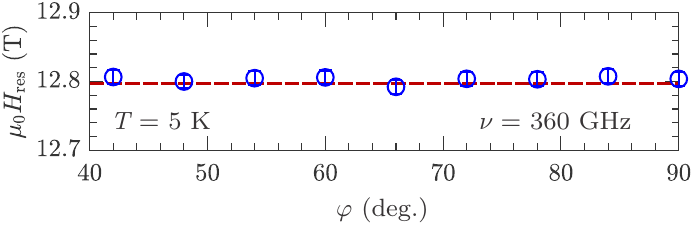}
 \caption{Angular dependence of the AFMR field for $\nu=360$~GHz and $T=5$~K. The dashed line corresponds to $g=2.01$. }\label{FIG:Adeps}
\end{figure}

The angular dependence of the magnetic resonance field in $\alpha$-MnTe exhibits the absence of detectable anisotropy, as Fig.~\ref{FIG:Adeps} demonstrates. At $360$~GHz,
the $g$-factor is invariant with respect to the magnetic-field direction in the $(0,0,1)$ plane. In order to extract the full width at half maximum $\Delta H$ accurately, we describe the ESR line with a Lorentian-based function that also accounts for possible profile distortions by a reactive magnetic susceptibility (see the Supplemental Material~\cite{SuppMatMnTe}). The resulting full-width at half-maximum at $5$~K is only $\mo\Delta H=50$~mT. The fluctuations of the resonance position are within a fraction of this value, setting the upper boundary on the relative in-plane $g$-factor variation as $\sim1.5\times10^{-3}$. This agrees with the high symmetry of $\alpha$-MnTe, and with the orbital momentum being $L=0$ for the ground state of Mn$^{2+}$ (hence, much reduced effects of the spin-orbit coupling). Our findings are also in line with the results of Ref.~\cite{RosslerGinga_arXiv_2025_MnTedomains}, confirming that the anisotropic, antiferromagnetic-domain-related behavior is limited to fields, much lower than $1$~T --- the minimal field where the resonance is detectable in our experimental frequency band.

\begin{figure}
 \includegraphics[width=0.45\textwidth]{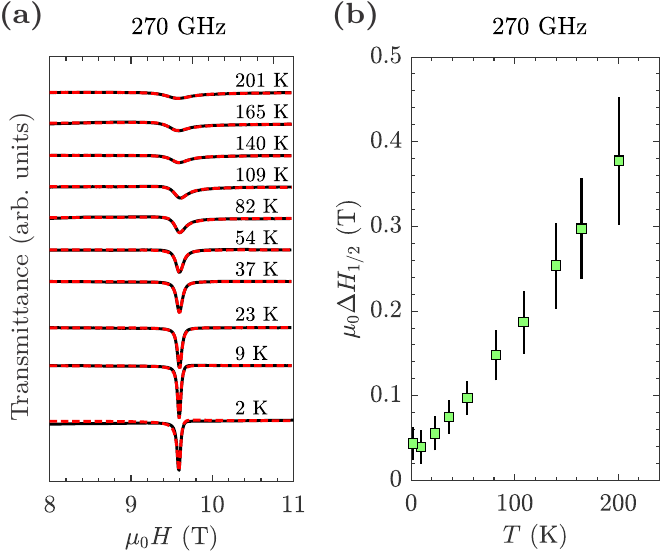}
 \caption{(a) Temperature dependence of ESR spectra at $\nu=270$~GHz ($\varphi=15^{\circ}$).
 Solid black lines correspond to experimental data, normalized to a value at maximal field. Dashed red lines correspond to Lorentian-based fits (see text). The spectra are offset for clarity. (b) Full width at half maximum of the magnetic resonance as the function of temperature ($\nu=270$~GHz).}\label{FIG:Tdeps}
\end{figure}

Upon warming, we observe substantial broadening of the resonance line as shown in Fig.~\ref{FIG:Tdeps}(a) for the frequency $270$~GHz. There is almost an order-of-magnitude rise of $\mo\Delta H$ from $50$~mT at low temperatures to $400$~mT near $200$~K, as the analysis results in Fig.~\ref{FIG:Tdeps}(b) reveal.

\begin{figure}
 \includegraphics[width=0.48\textwidth]{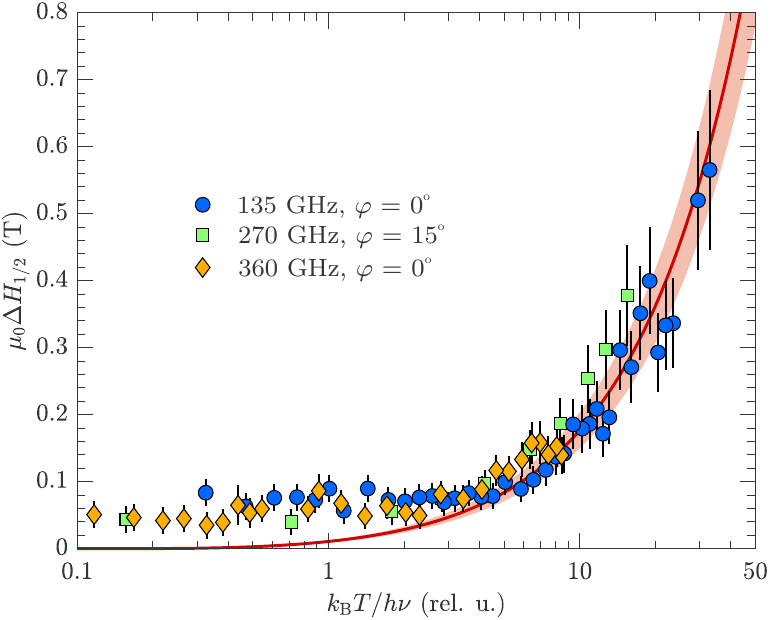}
 \caption{Resonance linewidth, obtained from spectra taken at $\nu=135$~GHz (blue circles, $\varphi=0^\circ$), $\nu=270$~GHz (green squares, $\varphi=15^\circ$), and $\nu=360$~GHz (orange diamonds, $\varphi=0^\circ$) as function of the normalized temperature $\kb T/h\nu$ (note the logarithmic scale). Solid lines and shading correspond to Eq.~(\ref{EQ:Bose0}) with $\Gamma_{0}/\kb=27(5)$~mK.}\label{FIG:FWHM}
\end{figure}

Remarkably, we find that the temperature-induced line broadening is strongly frequency-dependent and is less pronounced for larger frequencies.  As Fig.~\ref{FIG:FWHM} demonstrates, the data for three distinct frequencies ($135$, $270$, and $360$~GHz) form a single curve if plotted as the function of the ``reduced temperature'' $\kb T/h\nu$. Effectively, the linewidth is about $50$~mT for $\kb T \lesssim h\nu$ and grows along with $\kb T/h\nu$ for $\kb T \gtrsim h\nu$, representing, as we show below, a form of scaling behavior.

\textit{Discussion.}
Since the ground-state orbital momentum is absent in $L=0$ Mn$^{2+}$ ions, the broadening governed by the spin-orbit coupling~\cite{AbrahamBleany_2012_MagRes} is not relevant in $\alpha$-MnTe. This brings to the forefront the effects related to spin-spin interactions. In particular, the increase of the thermal magnon population upon the temperature growth should have an impact on the linewidth.

The experimentally found $\kb T/h\nu$ linewidth scaling is strongly suggesting magnon-magnon collisions~\cite{Dyson_PhysRev_1956_SpinWaveInteraction} as the important factor in antiferromagnetic relaxation. Indeed, the population of magnons is given by the Bose-Einstein distribution
\begin{equation}
 n_\mathbf{k}=\dfrac{1}{\exp(\frac{\hbar\omega_\mathbf{k}}{\kb T})-1},
 \label{EQ:Bose}
\end{equation}
where $\mathbf{k}$ is the quasi-momentum and $\hbar\omega_\mathbf{k}\simeq\sqrt{(g\mo\mub H)^2+(\hbar uk)^2}$ is the magnon dispersion of mode $\nu_1$ ($u$ is the spin-wave velocity). The ESR experiment probes magnons with $k\simeq0$, and their population $n_0=[\exp(\frac{g\mo\mub H}{\kb T})-1]^{-1}$ is the largest at any temperature.

We find that the available linewidth data follows the empirical dependence:
\begin{equation}\label{EQ:Bose0}
g\mo\mub \Delta H(T)=\Gamma_{0} n_{0}(T).\\
\end{equation}
Here the magnon population factor $n_0$ is governed by the Zeeman gap $\hbar\omega_{k=0}=g\mo\mub H=h\nu$. The observed $\kb T/h\nu$ behavior thus naturally follows from Eq.~(\ref{EQ:Bose0}), as $n_0=(\exp(\frac{h\nu}{\kb T})-1)^{-1}$. The parameter $\Gamma_{0}$ is the effective temperature-independent damping coefficient, representing the strength of interactions between the magnons.
With $\Gamma_{0}/\kb=27(5)$~mK the experimental data above $30$~K can be described very well for the three frequencies we used.

While this framework provides a good description of the data, the other mechanisms could also be partially responsible for the observed AFMR line broadening. In particular, in the case of $\alpha$-MnTe the magnons can also scatter on itinerant electrons, on crystal imperfections, and on phonons. The influence of the extremely diluted electron population is estimated to be negligible in $\alpha$-MnTe. The contribution of impurities is temperature independent and would be most relevant in the low-$T$ limit, as we see below. The contribution of phonons, on the other hand, may be noticeable at high temperatures, but in the case of $\alpha$-MnTe there are reasons to expect it to be quite suppressed as well. We discuss the potential contributions of electrons and phonons  in more detail in the Supplemental Material~\cite{SuppMatMnTe} (see also Refs.~\cite{MagninDiep_PRB_2012_MnTeresist,LutovinovReizer_SovPhysJETP_1979_FMmetalRelax,HeZhang_JMST_2011_MnTeresist,Orbach_RoyalProc_1961_SpinlatticeRelaxation,KaganovTsukernik_SovPhysJETP_1959_MagnPhon,AkhiezerBariakhtarPeletminskii_1968_SpinWaves,StreibVidalSilva_PRB_2019_MagPhscatteringYIG}). The defects would be relevant only at the lowest temperatures, and we discuss their role towards the end of this section.

Thus, we argue that Eq.~(\ref{EQ:Bose0}) describes the most important relaxation mechanism at the temperatures we deal with, in line with considerations for the other antiferromagnetic materials~\cite{HarrisKumar_PRB_1971_SpinWaveHydrodynamics,RezendeWhite_PRB_1976_AFMRrelaxation,BayrakciKeller_Science_2006_MnF2lifetimes}.
Such direct proportionality of damping to the long-wavelength magnon population is not unprecedented. Similar behavior has been theoretically predicted~\cite{DamleSachdev_PRB_1998_1Dgapped} and experimentally observed in one-dimensional gapped magnets~\cite{NafradiKeller_PRL_2011_IPAmagnon,Schmidiger_2014_PhDthesis}. The physical picture behind this is straightforward: the lifetime of the quasiparticle is limited by the collision probability, which is in turn defined by the number of quasiparticles already present.
Similar to those studies, in $\alpha$-MnTe we deal with the low-temperature limit. Most of the data corresponds to $T/T_\mathrm{N}$ between $0.1$ and $0.3$. Other studies on AFMR relaxation in ordered three-dimensional antiferromagnetic systems consider wider temperature ranges, making large momenta $k$, and collisions involving larger numbers of magnons more relevant. This leads to more complex temperature dependencies~\cite{HarrisKumar_PRB_1971_SpinWaveHydrodynamics,RezendeWhite_PRB_1976_AFMRrelaxation}. For similar reasons, we also can neglect possible temperature-induced changes of magnon parameters (e.g. spin-wave velocity). All such factors would be most pronounced in the vicinity of $T_\mathrm{N}$, where the damping is dominated by the critical fluctuations~\cite{ForsterGarcia_PRB_2013_Pb2VOP2O8ESR,BennerBoucher_1990_CriticalMagneticResonance}. For the given temperature range such fluctuations, whose effect is frequency independent~\cite{BennerBoucher_1990_CriticalMagneticResonance}, should not be the leading factor.

We observe a small constant linewidth of about $50$~mT (equivalent to $\sim 70$~mK) below the temperature of nearly $30$~K. This suggests that some other, temperature-independent factor becomes more important than the magnon collisions. We attribute this relaxation mechanism as the scattering of magnons on ubiquitous crystal defects.
In the present case, effective $\sim 70$~mK residual damping is only $\sim 10^{-3}$ of the leading exchange interaction $J_{c}/\kb=46.3$~K. This is a fine effect in $\alpha$-MnTe. Some further details are discussed in Supplemental Material~\cite{SuppMatMnTe} (see also Refs.~\cite{ZhitomirskyChernyshev_RMP_2013_DecayReview,ChernyshevZhitomirsky_PRL_2012_BNPOspinecho,ShpyrkoIsaacs_Nature_2007_AntiferromagneticDomians,PetschShen_PRB_2025_impurityrelaxation}).

\textit{Conclusions.}
Summarizing, we studied the low-energy spin dynamics of altermagnetic $\alpha$-MnTe by means of high-field AFMR, with the magnetic field applied along the easy anisotropy plane. We have tracked the frequency-field, angular, and temperature dependencies of the pseudo-Goldstone mode below $T_\mathrm{N}$. The detected low-frequency AFMR mode has $g_\mathrm{eff} = 2.01$ which (together with the observed isotropic AFMR field behavior) agrees with the quenched orbital momenta for Mn$^{2+}$ ions. At low temperatures, the mode is remarkably narrow, but exhibits substantial broadening with increasing temperature. We interpret such a behavior in terms of the magnon-magnon scattering.

\textit{Acknowledgments.}
This work was supported by the Deutsche Forschungsgemeinschaft through the W\"{u}rzburg-Dresden Cluster of Excellence $ctd.qmat$ --- Complexity, Topology and Dynamics in Quantum Matter (EXC 2147, Project No.\ 390858490) and the SFB 1143 (Project No.\ 247310070), as well as by HLD at HZDR, member of the European Magnetic Field Laboratory (EMFL). We thank V. Hasse for technical assistance with the crystal growth, and M. Uhlarz for help with the Laue-diffraction crystal characterization.

\bibliography{MnTeBib.bib}

 \foreach \x in {1,...,\numbersupplementpages}
    {
        \clearpage
        \includepdf[pages={\x}]{SuppMat_MnTe.pdf}
    }

\end{document}